\documentclass[]{article}
\usepackage{xcolor}
\usepackage{gensymb}

\input{macros.sty}

\makeatletter
\def\paragraph{\@startsection{paragraph}{4}%
  \z@\z@{-\fontdimen2\font}%
  {\normalfont\bfseries}}
\makeatother

\usepackage{graphicx}
\usepackage{newtxtext}
\usepackage{newtxmath}
\usepackage{natbib}
\usepackage{hyperref}
\hypersetup{
    colorlinks = true,
    urlcolor   = blue,
    citecolor  = black,
}
\usepackage{authblk}

\newtheorem{theorem}{Theorem}
\newcommand{\RomanNumeralCaps}[1]
\linenumbers

\title{Tensor-based flow reconstruction from optimally located sensor measurements}
\author{ Mohammad Farazmand and Arvind K. Saibaba
}

\affil{Department of Mathematics, North Carolina State University, 2311 Stinson Dr., Raleigh, NC 27695}
\date{}

\begin{document}
\maketitle

\begin{abstract}
Reconstructing high-resolution flow fields from sparse measurements is a major challenge in fluid dynamics. Existing methods often vectorize the flow by stacking different spatial directions on top of each other, hence \MF{confounding} the information encoded in different dimensions. 
Here, we introduce a tensor-based sensor placement and flow reconstruction method which retains and exploits the inherent multidimensionality of the flow. We derive estimates for the flow reconstruction error, storage requirements and computational cost of our method. We show, with examples, that our tensor-based method is significantly more accurate than similar vectorized methods. Furthermore, the variance of the error is smaller when using our tensor-based method. While the computational cost of our method is comparable to similar vectorized methods, it reduces the storage cost by several orders of magnitude. The reduced storage cost becomes even more pronounced as the dimension of the flow increases. We demonstrate the efficacy of our method on \AS{three examples: a chaotic Kolmogorov flow, in-situ and satellite measurements of the global sea surface temperature, and 3D unsteady simulated flow around a marine research vessel.} 
\end{abstract}

\section{Introduction}
Numerical simulations of fluid flows are carried out with ever growing spatial resolution. In contrast, observational data is limited to relatively coarse sensor measurements. This dichotomy inhibits efficient integration of experimental data with existing high-fidelity computational methods to enable detailed and accurate flow analysis or prediction.

As we review in \S\ref{sec:RelWork}, several methods have been developed to address this disconnect. In particular, flow reconstruction methods seek to leverage offline high-resolution simulations to estimate the entire flow field from coarse online observations. All existing methods vectorize the simulation data by stacking different spatial dimensions on top of each other. Vectorization is convenient since it enables one to use familiar linear algebra techniques. However, this approach inevitably leads to loss of information encoded in the inherent multidimensional structure of the flow.

Here, we propose a tensor-based sensor placement and flow reconstruction method which retains and exploits the multidimensional structure of the flow.
Our method significantly increases the accuracy of the reconstruction compared to similar vectorized methods. We quantify this accuracy by deriving an upper bound for the reconstruction error and show that the resulting approximation exactly interpolates the flow at the sensor locations (assuming the measurements are noise-free). Additionally, the proposed method has a smaller memory footprint compared to the similar vectorized methods, and is scalable to large datasets using randomized techniques.  We demonstrate the efficacy of our method on \AS{three examples: direct numerical simulation of a turbulent Kolmogorov flow, in situ and satellite sea surface temperature (SST) data, and 3D unsteady simulated flow around a marine research vessel.} \MF{We emphasize that our focus here is only on flow reconstruction and not temporal prediction; nonetheless, the reconstructed flow can be subsequently used as input to high-fidelity or reduced-order predictive models.}

\subsection{Related work}\label{sec:RelWork}
Due to the broad applications of flow estimation from sparse measurements, there is an expansive body of work on this subject (see~\cite{Callaham2019} for a thorough review). 
Here, we focus on the so-called library-based methods. These methods seek to reconstruct the flow field by leveraging the sparse observational measurements to interpolate a pre-computed data library comprising high-fidelity numerical simulations. 
More specifically, consider a scalar quantity $g(\M x,t)$ which we would like to reconstruct from its spatially sparse measurements. 
For instance, this quantity may be a velocity component, a vorticity component, pressure, or temperature.
The data library is a matrix $\Phi\in \mathbb R^{N\times T}$ whose columns are formed from vectorized high-resolution simulations. Here, $N$ denotes the number of collocation points used in the simulations. 
The columns of $\Phi$ may coincide with the quantity of interest $g$ or be derived from this quantity, e.g., through proper orthogonal decomposition (POD) or dynamic mode decomposition (DMD). The observational data $\M y\in \mathbb R^r$ is a vector containing $r$ measurements of the quantity $g$ at a particular time. Library-based methods seek to find a map $F:\mathbb R^{N\times T}\times \mathbb R^r\to \mathbb R^N$ such that $\M g \simeq F(\Phi,\M y)$. Here, $\M g\in\mathbb R^N$ is a vector obtained by stacking the quantity of interest $g(\M x,t)$ at the collocation points.

Library-based methods differ in their choice of the data matrix $\Phi$ and the methodology for finding the map $F$.
A common choice for the columns of the data matrix is the POD modes~\citep{Willcox2004,Willcox2006}, although DMD modes~\citep{Nabi2017,Zhang2021} and flow snapshots~\citep{Brunton2021} have also been used. \cite{Willcox2004} use the gappy POD algorithm of~\cite{Sirovich1995} to reconstruct the flow. They obtain the map $F$ by solving a least squares problem which seeks to minimize the discrepancy between the observations $\M y$ and the reconstructed flow $F(\Phi,\M y)$ at the sensor locations (also see~\cite{Willcox2006}).  

Discrete Empirical Interpolation Method (DEIM) takes a similar approach, but the map $F$ is a suitable oblique projection on the linear subspace spanned by the columns of $\Phi$. DEIM was first developed by~\cite{Sorensen2010} for efficient reduced-order modeling of nonlinear systems and was later used for flow reconstruction~\citep{Drmac2016,Lin2021}. Several subsequent modifications to DEIM have been proposed, e.g., to lower its computational cost~\citep{Willcox2014} and to generalize it for use with weighted inner products~\citep{Drmac2018}.  

It is well-known that both gappy POD and DEIM suffer from overfitting~\citep{Drmac2020}. Consequently, if the sensor data $\M y$ is corrupted by significant observational noise, the reconstruction error will be large. \cite{Callaham2019} use sparsity promoting techniques from image recognition to overcome this problem (also see~\cite{Chu2021}). Their reconstruction map $F$ is obtained by solving a sparsity-promoting optimization problem with the constraint that the reconstruction error is below a prescribed threshold. The resulting method is robust to observational noise. However, unlike gappy POD and DEIM, the reconstruction map $F$ cannot be expressed explicitly in terms of the training data $\Phi$.

Yet another flow reconstruction method is to represent the map $F$ with a neural network which is trained using the observations $\M y$ and the data matrix $\Phi$. For instance, \cite{Hesthaven2019} use an autoencoder to represent the reconstruction map $F$ (also see~\cite{Taira2019,Jameson2019,Erichson2020}). \MF{Unlike DEIM, where the reconstruction map $F$ is a linear combination of modes, neural networks can construct nonlinear maps from the observations $\M y$ and the library $\Phi$}. These machine learning methods have shown great promise; however, the resulting reconstructions are not explicit, or even interpretable, since they are only available as a complex neural network.

\MF{With the notable exception of convolutional neural networks~\citep{Taira2019,Jameson2019}, almost all existing methods treat the data as a vector by stacking different spatial dimensions on top of each other.} This inevitably leads to loss of information encoded in the inherent multidimensionality of the flow. Here, we propose a tensor-based method which retains and exploits this multidimensional structure. Our method is similar to the tensor-based DEIM which was recently proposed by~\cite{kirsten2022multilinear} for model reduction; but we use it for flow reconstructions which is the focus of this paper. Numerical experiments show that the resulting reconstructions are more accurate compared to similar vectorized methods, because of our method's ability to capture and exploit the inherent multidimensional nature of the data. The computational cost of our tensor-based method is comparable to the vectorized methods and can be further accelerated using randomized methods. Furthermore, the tensor-based method requires much less storage compared to the vectorized methods. This is especially important in large-scale 3D flows where the storage costs can be substantial~\citep{Gelss2019}.
Although our method is a tensorized version of DEIM, a similar tensor-based approach can be applied to other flow reconstruction methods such as gappy POD, sparsity-promoting methods, and autoencoders.



\section{Tensor-based flow reconstruction}
\subsection{Set-up and preliminaries}
Let $g(\M x,t)$ denote the quantity of interest at time $t$ that we would like to reconstruct. 
This quantity may for instance be a velocity component, a vorticity component, pressure, or temperature. 
The spatial variable is denoted by $\M x\in \Omega\subset \mathbb R^d$, where $\Omega$ is the flow domain with $d=2$ or $d=3$ for two- and three-dimensional flows, respectively. 
In numerical simulations, the quantity of interest $g$ is discretized on a spatial grid of size $N_1\times N_2\times\cdots\times N_d$ and saved as a tensor $\T G\in\mathbb R^{N_1\times\cdots\times N_d}$. 

We denote entries of $\T G$ by $g_{i_1,\dots,i_d}$ where $1 \leq i_n \leq N_n$ and $1 \leq n \leq d$. There are $d$ different matrix unfoldings of $\T G$, also called {\em matricizations}, which we denote by $\M{G}_{(n)} \in \R^{N_n \times (\prod_{j \neq n} N_j)}$. The mode-$n$ product of a tensor $\T{G}$ with a matrix $\M{M} \in \R^{R \times N_n}$ is denoted as $\T{Y} = \T{G} \times_n \M{M}$ with entries  
\begin{equation}
    y_{i_i,\dots, j,\dots, i_N} = \sum_{k=1}^R g_{i_1,\dots,k,\dots,i_N}m_{jk} \qquad 1 \leq j \leq R.
\end{equation} 

In terms of matrix unfoldings, it can be expressed as $\M{Y}_{(n)} = \M{M} \M{G}_{(n)}$. For matrices $\M{A}$ and $\M{B}$ of compatible dimensions $\T{G} \times_m \M{A} \times_n \M{B} = \T{G}  \times_n \M{B} \times_m \M{A}$ if $m \neq n$ and $\T{G} \times_n \M{A} \times_n \M{B} = \T{G} \times_n \M{BA}$. Associated with every tensor $\T{G}$ is a multirank $(R_1,\dots,R_d)$ where $R_n = \text{rank}(\M{G}_{(n)})$. The Frobenius norm of a tensor is $\|\T{G}\|_F^2 = \sum_{i_1,\dots,i_d}g_{i_1,\dots,i_d}^2$.  We refer to~\cite{kolda2009tensor} for a detailed review of tensor operations. 

\subsection{Vectorized POD-DEIM} 
We first review the vectorized form of POD-DEIM from which our tensor-based method is derived. We refer to this method as vector-DEIM, for short.
In vector-DEIM approach for sensor placement~\citep{manohar2018data,clark2018greedy}, the training data is constructed as the snapshot matrix, $ \M{G} = \bmat{ \M{g}_1 & \dots & \M{g}_T } \in \R^{N \times T}$, 
where each column $\M{g}_j \in \R^N$ represents a vectorized snapshot of the quantity of interest $g(\M x,t_j)$. We assume that the vectors are centered, which means that the mean is subtracted from each column. We would like to pick $r \leq \min\{N,T\}$ number of sensor locations at which to collect data (see Fig.~\ref{fig:schem}).

We first compute the truncated singular value decomposition (SVD) $\M{G} \approx \M{U}_r \M\Sigma_r \M{V}_r^\top$, 
where the columns of $\M{U}_r\in\mathbb R^{N\times r}$ coincide with the POD modes. In DEIM, we compute the column pivoted QR factorization~\cite[Section 5.4.2]{GaVL} of $\M{U}_r$; that is, we compute $\M{U}_r^\top \bmat{\M{P}_1 & \M{P}_2 } = \M{Q}_1 \bmat{ \M{R}_{11} & \M{R}_{12}}$, 
where $\M{Q}_1 \in \R^{r\times r}$ is orthogonal, $\M{R}_{11} \in \R^{r\times r}$ is upper triangular, and $\M{R}_{12} \in \R^{r\times N}$. The matrix $\M{P} = \bmat{\M{P}_1 & \M{P}_2 }$ is a permutation matrix.  Suppose we have $\M{P}_1 = \bmat{\M{I}(:,{i_1}) & \dots & \M{I}(:,i_r)}$, where $\M I$ is the $N\times N$ identity matrix.
Then, the matrix $\M{P}_1 \in \R^{N\times r}$ contains columns from the identity matrix indexed by the set $\mc{I} = \{i_1,\dots, i_r\}$. The spatial locations corresponding to the index set $\mc{I}$ are used as the optimal sensor locations, which may not be uniquely determined. 

Suppose we want to reconstruct the flow field corresponding to the vector $\V{f} \in \R^N$. We collect measurements at the indices corresponding to $\mc{I}$, given by the vector $\V{f}(\mc{I})$. In other words, $\V{f}(\mc{I})$ is the available sensor measurements of the vector $\V{f}$.
To reconstruct the full flow field $\V{f}$, we use the approximation
\begin{equation}\label{eqn:vecrecon}
    \V{f} \approx \M{U}_r \M{U}_r(\mc{I},:)^{-1}\V{f}(\mc{I}) = \M{U}_r(\M{S}^\top\M{U}_r)^{-1}\M{S}^\top\V{f},
\end{equation}
where $\M{S} =\M{P}_1 $. 
This approximation provides insight into how the indices $\mc{I}$ should be selected. Since $\M{U}_r$ has rank $r$, it is guaranteed to have $r$ linearly independent rows and $\M{S}^\top\M{P}_1$ is invertible.  The index set $\mc{I}$ is chosen in such a way that the corresponding rows of $\M{U}_r$ are well-conditioned.  
The error in the training set takes the form
 \begin{equation}\label{eqn:poderror} 
 \begin{aligned}\| \M{G} - \M{U}_r \M{U}_r(\mc{I},:)^{-1} \M{G}(\mc{I},:)\|_F \leq & \>  \|( \M{S}^\top
 \M{U}_r)^{-1}\|_2 \left(\sum_{j=r+1}^{\min\{N,T\}}\sigma_{j}^2 (\M{G}) \right)^{1/2 },
\end{aligned}\end{equation}
where $\sigma_j(\M{G})$ represents the singular values of $\M{G}$. 
In the above expression for the error, $ \left(\sum_{j=r+1}^{\min\{N,T\}}\sigma_{j}^2 (\M{G}) \right)^{1/2} = \|\M{G} - \M{U}_r\M{U}_r^\top\M{G}\|_F$ represents the error in the POD approximation due to the truncated singular values, which is amplified by the factor $\|( \M{S}^\top
\M{U}_r)^{-1}\|_2$ which arises due to the DEIM approximation. The error in the test dataset can be obtained using Lemma 3.2 of~\cite{Sorensen2010}. 
 
 \subsection{Tensor-based POD-DEIM} 
\begin{figure}
\centering
\includegraphics[width=\textwidth]{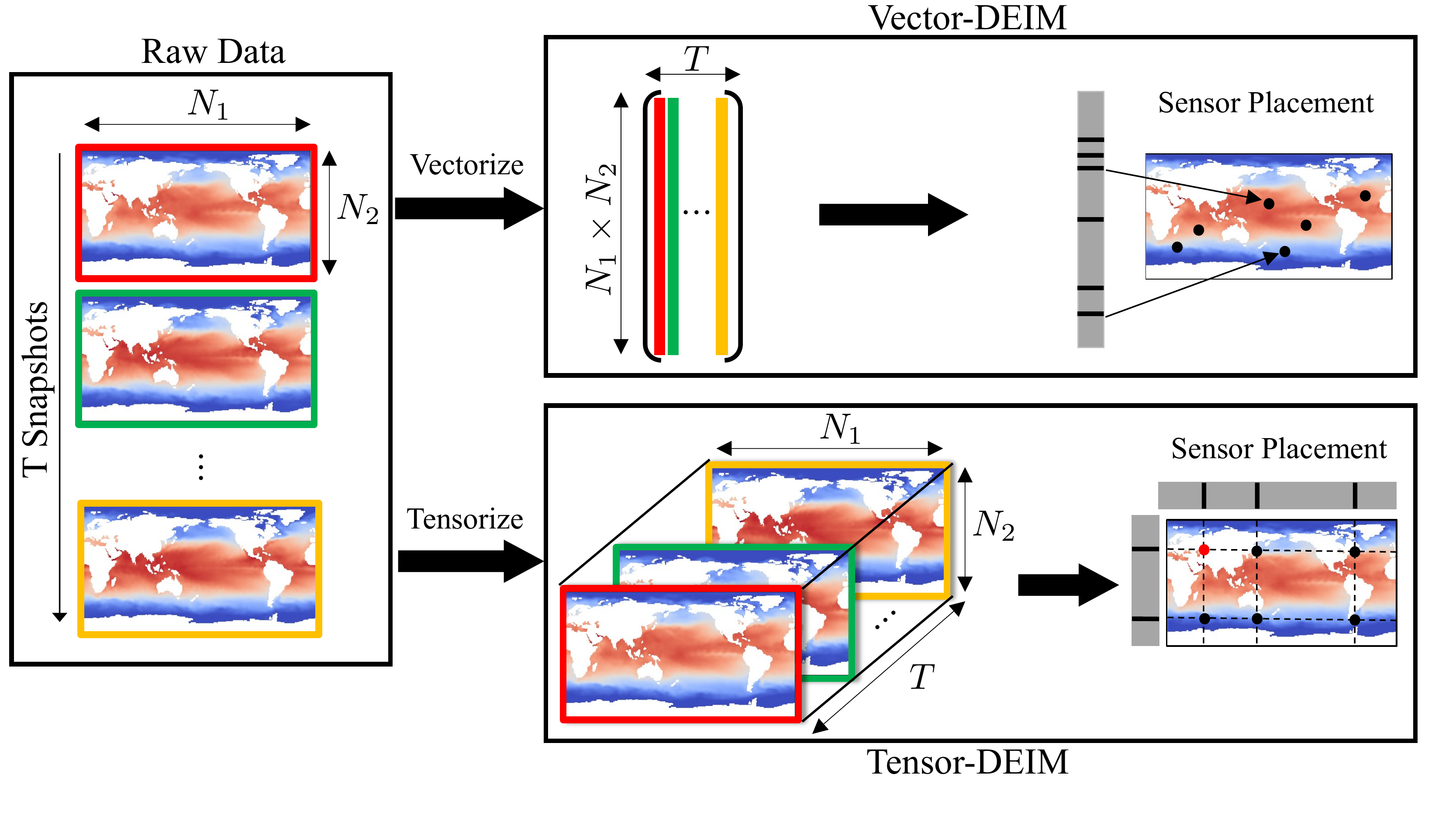}
\caption{\AS{Schematic representation of the workflow in vector-DEIM and tensor-DEIM. The horizontal and vertical black bars mark the entries that are used for sensor placement (black circles). In tensor-DEIM, some sensors may fall on land and will be discarded (e.g., the one marked by the red circle).}}
\label{fig:schem}
\end{figure}

 In vector-DEIM, the snapshots are treated as vectors meaning that the inherent multidimensional structure of the flow is lost. In order to fully exploit this multidimensional structure, we use tensor-based methods. We refer to the resulting tensorized version of POD-DEIM as tensor-DEIM, for short. We consider the collection of snapshots in the form of a tensor $\T{G} \in \R^{N_1 \times \dots \times N_d \times T }$ of order $d+1$, where $d$ represents the number of spatial dimensions and $\prod_{j=1}^dN_j = N $ is the total number of grid points (see Fig.~\ref{fig:schem}). With this notation, the snapshot matrix $\M{G}$ in vector-DEIM can be expressed as $\M{G}_{(d+1)}^\top$; that is the transpose of the mode-$(d+1)$ unfolding. 

Suppose we wanted to collect data at $r = \prod_{n=1}^d r_n$ sensor locations. In tensor-DEIM, we first compute the truncated SVD of the first $d$ mode unfoldings.  That is, we compute $\M{G}_{(n)} \approx \M{U}_r^{(n)} \M\Sigma_r^{(n)} (\M{V}_r^{(n)})^\top $ where $\M{U}_r^{(n)} \in \R^{N_n \times r_n}$. For ease of notation, we define $\M\Phi_n := \M{U}_r^{(n)}$. Next, we compute the column-pivoted QR factorization of $\M\Phi_n$ as 
\begin{equation}
 \M\Phi_n^\top\bmat{\M{P}_1^{(n)} & \M{P}_2^{(n)} } = \M{Q}_{1}^{(n)} \bmat{\M{R}_{11}^{(n)} & \M{R}_{12}^{(n)}} \qquad 1 \leq n \leq d.    
\end{equation}
Here $\bmat{\M{P}_1^{(n)} & \M{P}_2^{(n)} }$ is a permutation matrix. Once again, for ease of notation, we set $\M{S}_n := \M{P}_1^{(n)}$; this matrix contains the columns from the $N_n \times N_n$ identity matrix corresponding to the indices $\mc{I}_n = \{i_1^{(n)}, \dots, i_{r_n}^{(n)}\}$. Then the sensors can be placed at the spatial locations corresponding to the index set $\mc{I}_1 \times \dots \times \mc{I}_d$. Note that when $d=1$ tensor-DEIM reduces to vector-DEIM.
 
 Given a new data $\T{F} \in \R^{N_1 \times \dots \times N_d}$, such that $\V{f} = \text{vec}(\T{F})$, we only need to collect data at the indices corresponding to $\mc{I}_1 \times \dots \times \mc{I}_d$, that is we measure $\T{F}(\mc{I}_1,\dots,\mc{I}_d)$. To reconstruct the flow field from these measurements, we compute 
 \begin{equation}
     \T{F} \approx  \T{F}(\mc{I}_1,\dots,\mc{I}_d) \times_{n=1}^d \M\Phi_n (\M{S}_n^\top\M\Phi_n)^{-1} = \T{F} \times_{n=1}^d \M\Phi_n (\M{S}_n^\top\M\Phi_n)^{-1}\M{S}_n^\top.
 \end{equation}
 The second expression, while equivalent to the first, is more convenient for the forthcoming error analysis. The approximation just derived satisfies the interpolation property; that is, the approximation exactly matches the function $\T{F}$ at the sensor locations assuming the measurements are noise-free. To see this, denote $\T{F}_{\rm TDEIM} := \T{F} \times_{n=1}^d \M\Phi_n (\M{S}_n^\top\M\Phi_n)^{-1}\M{S}_n^\top. $ Then 
 \begin{equation}
 \begin{aligned} \T{F}_{\rm TDEIM}(\mc{I}_1,\dots,\mc{I}_d) = & \> \T{F}_{\rm TDEIM } \times_{n=1}^d \M{S}_n^\top =   \T{F} \times_{n=1}^d \M{S}_n^\top \M\Phi_n (\M{S}_n^\top\M\Phi_n)^{-1}\M{S}_n^\top  \\
 = & \>\T{F} \times_{n=1}^d \M{S}_n^\top  = \T{F}(\mc{I}_1,\dots,\mc{I}_d).
 \end{aligned}
  \end{equation}

 The following theorem provides an expression for the error in the approximation as applied to the training dataset.
 \begin{theorem} \label{thm:mainten} 
 Suppose $\M\Phi_n \in \R^{N_n\times r_n}$ is computed from the truncated rank-$r_n$ SVD of the mode unfolding $\M{G}_{(n)}$ and $\M{S}_n$ is obtained by computing column pivoted QR of $\M\Phi_n^\top$ such that $\M{S}_n^\top\M\Phi_n$ is invertible for $1 \leq n \leq d$. Define $\M\Pi_n := \M\Phi_n (\M{S}_n^\top\M\Phi_n)^{-1}\M{S}_n^\top$ and  assume $1 \leq r_n < N_n$ for $1 \leq n \leq d$. 
 Then 
\begin{equation}
\| \T{G} - \T{G} \times_{n=1}^d \M\Pi_n\|_F  \leq  \left(\prod_{ n=1}^d \|(\M{S}_n^\top\M\Phi_n)^{-1} \|_2  \right)  \left(\sum_{n=1}^d \sum_{k>r_n} \sigma_{k}^2(\M{G}_{(n)}) \right)^{1/2}.
\label{eq:err_tensor}
\end{equation} 
\end{theorem}
The proof of this theorem is given in Appendix~\ref{sec:app}. The interpretation of this theorem is as follows: the term $\left(\sum_{n=1}^d \sum_{k>r_n} \sigma_{k}^2(\M{G}_{(n)}) \right)^{1/2}$ represents the error due to the truncated SVD in each mode, and $\left(\prod_{ n=1}^d \|(\M{S}_n^\top\M\Phi_n)^{-1} \|_2  \right)$ represents the amplification due to the selection operator across each mode. The upper bound~\eqref{eq:err_tensor} is similar to the upper bound~\eqref{eqn:poderror} for vector-DEIM. However, it is difficult to establish which bound is tighter {\em a priori}. \MF{As will be shown} in \S\ref{sec:results}, numerical evidence strongly suggests that the error due to tensor-DEIM is much lower than vector-DEIM.

The error in the test sample can be determined using Proposition 1 from~\cite{kirsten2022multilinear}, which gives 
\begin{equation}\label{eqn:tdeim_test_err}
\| \T{F} - \T{F} \times_{n=1}^d \M\Pi_n\|_F  \leq    \left(\prod_{n=1}^d \| (\M{S}_n^\top\M\Phi_n)^{-1}\|_2  \right) \|\T{F}  - \T{F} \times_{n=1}^d \M\Phi_n\M\Phi_n^\top\|_F.
\end{equation} 
If strong rank-revealing QR (sRRQR) algorithm~\cite[Algorithm 4]{GuEis} with parameter $f = 2$ is used to compute the selection operators $\M{S}_n$, then the bound in Theorem~\ref{thm:mainten} simplifies to 
\begin{equation}
    \| \T{G} - \T{G} \times_{n=1}^d \M\Pi_n\|_F  \leq    \left(\prod_{n=1}^d \sqrt{1 + 4r_n (N_n-r_n)} \right) \left(\sum_{n=1}^d  \sum_{k>r_n} \sigma_{k}^2(\M{G}_{(n)}) \right)^{1/2}.
\end{equation} 
See~\cite[Lemma 2.1]{Drmac2018} for details. 

\subsubsection{Storage cost}
We only need to store the bases $\M\Phi_n \in \R^{N_n\times r_d}$, which costs $\sum_{n=1}^d N_nr_n$ entries. Compare this with vector-DEIM which requires $  r \prod_{n=1}^d N_n = rN$ entries. Assuming $N_1 = \dots = N_d $ and $r_1 = \dots =  r_d$, the ratio of storage cost of tensor-DEIM to that of vector-DEIM is 
\begin{equation}
    \text{ratio}_\text{stor} := \frac{\sum_{n=1}^d r_nN_n}{rN} = \frac{dr_1 N_1}{r_1^dN_1^d} = \frac{d}{r_1^{d-1}N_1^{d-1}}.
\end{equation} 
Therefore, the compression available using tensors can be substantial when the dimension $d$, the grid size $N_1,$ and/or the number of sensors $r_1$ are large. As an illustration in three spatial dimensions, $d=3$, let $N_1 = N_2 = N_3 = 1024$ grid points, and the target rank $r_1 = r_2 = r_3 = 25$; the fraction of the storage cost of tensor-DEIM bases, compared to vector-DEIM, is $3/(25^2\cdot 1024^2) \times 100\% \approx 4.6\times 10^{-7} \%$. \MF{Similar savings in terms of storage costs were also reported in~\cite{Gelss2019} who used tensor-based methods for data-driven discovery of governing equations.}
 
 \subsubsection{Computational Cost} 
 The cost of vector-DEIM is essentially the cost of computing an SVD on a $N\times T$ matrix, which is $\mc{O}(NT^2)$ floating point operations (flops) assuming $T \leq N$. The cost of computing the tensor-DEIM bases is $\mc{O}(T\prod_{j=1}^dN_j \sum_{n=1}^dN_n)$ flops. Tensor-DEIM is slightly more expensive since it has to compute $d$ different SVDs compared to vector-DEIM. This computational cost can be amortized by using the sequentially truncated higher-order SVD of~\cite{vannieuwenhoven2012new}. The cost of computing the indices that determine the sensor locations is $\mc{O}(\sum_{n=1}^d N_nr_n^2)$ flops for tensor-DEIM which is much cheaper than vector-DEIM which costs $\mc{O}(N r^2)$ flops. 
 
When the training dataset is large, computing the truncated SVD approximation can be expensive. One way to accelerate this computation is to use randomized methods as in~\citep{minster2020randomized}. Suppose the randomized higher order SVD is used, then the computational cost is $\mc{O}(T \prod_{j=1}^dN_j \sum_{n=1}^dr_n) $  flops. This cost is substantially less than the cost of both vector-DEIM and tensor-DEIM. 
 
\section{Results and discussion}\label{sec:results}
In the numerical experiments, we use QR with column pivoting as implemented in MATLAB for both vector-DEIM and tensor-DEIM.
\subsection{Kolmogorov flow}
Kolmogorov flow refers to a turbulent flow with periodic boundary conditions and a sinusoidal forcing. Here we consider the two-dimensional Kolmogorov flow, 
\begin{equation}
    \partial_t \omega + \mathbf u\cdot\nabla \omega = \nu \Delta \omega - n\cos (ny),
    \label{eq:kolm}
\end{equation}
where $\mathbf u= (\partial_y \psi,-\partial_x\psi)$ is the fluid velocity field with the stream function $\psi(x,y,t)$ and $\omega= -\Delta \psi$ is the vorticity field.
We consider a two-dimensional domain $\M x=(x,y)\in [0,2\pi]\times [0,2\pi]$ with periodic boundary conditions. The forcing wavenumber is $n=4$ and $\nu = \mbox{Re}^{-1}$ is the inverse of the Reynolds number $\mbox{Re}$.
\begin{figure}
    \centering
    \includegraphics[width=\textwidth]{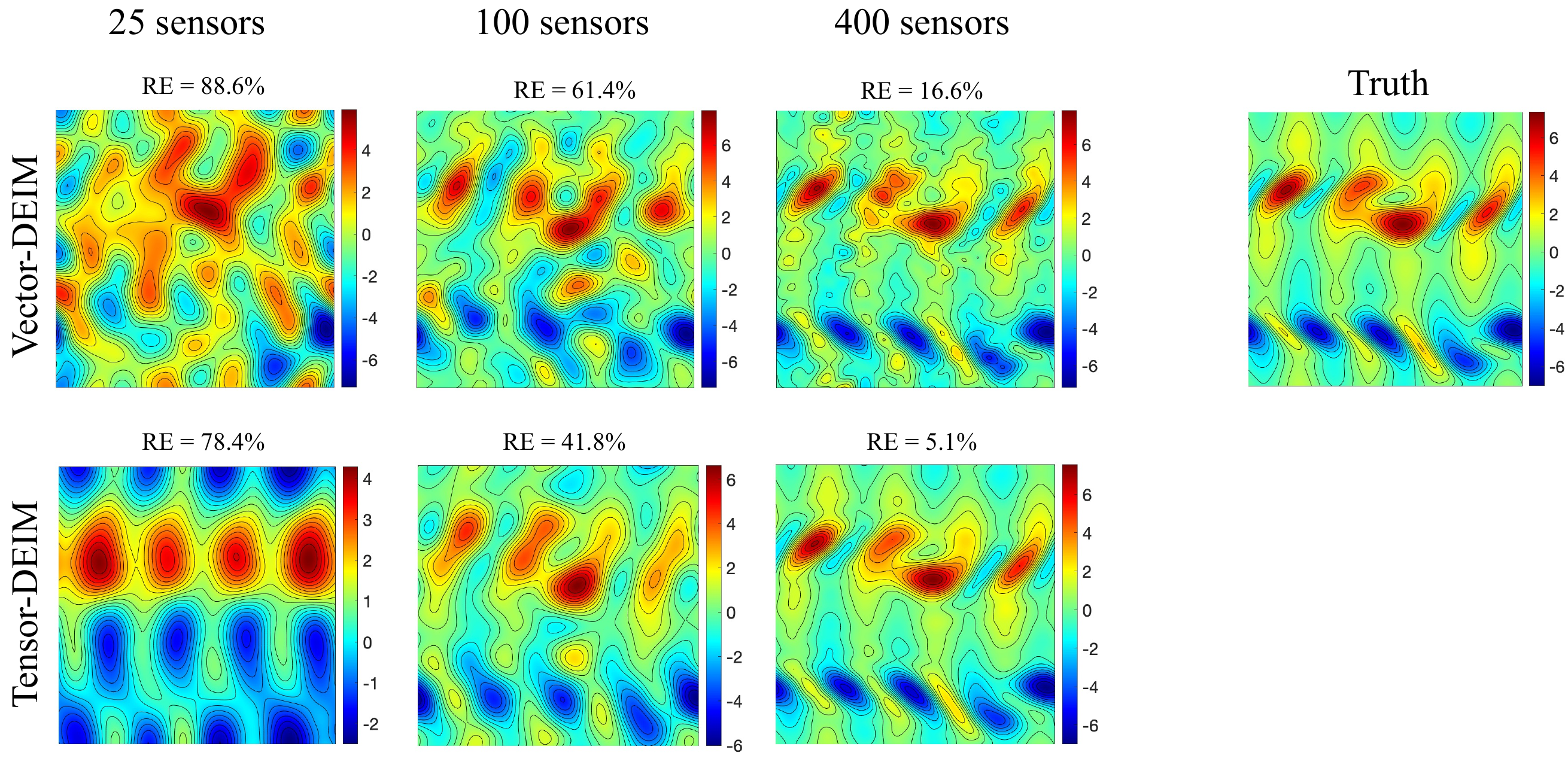}
    \caption{Comparing vector-DEIM and tensor-DEIM on the Kolmogorov flow. RE denotes the relative error of the reconstruction.}
    \label{fig:kolm_w}
\end{figure}

We numerically solve equation~\eqref{eq:kolm} using a standard pseudo-spectral method with $128\times 128$ modes and $2/3$ dealiasing. 
The temporal integration is carried out with the embedded Runge--Kutta scheme of~\cite{RK45}. 
The initial condition is random and is evolved long enough to ensure that the initial transients have died out before any data collection is performed. Then $10^3$ vorticity snapshots are saved, each $\Delta t=5$ time units apart. \MF{The time increment $\Delta t$ is approximately 10 times the eddy turnover time $\tau_e\simeq 0.5$ of the flow, ensuring that the snapshots are not strongly correlated.}
First $75\%$ of the data are used for training. The remaining 25\% are used for testing. 
The training data forms the data tensor $\T G\in\mathbb R^{128\times 128\times 750}$.

\begin{figure}
    \centering
    \includegraphics[width=\textwidth]{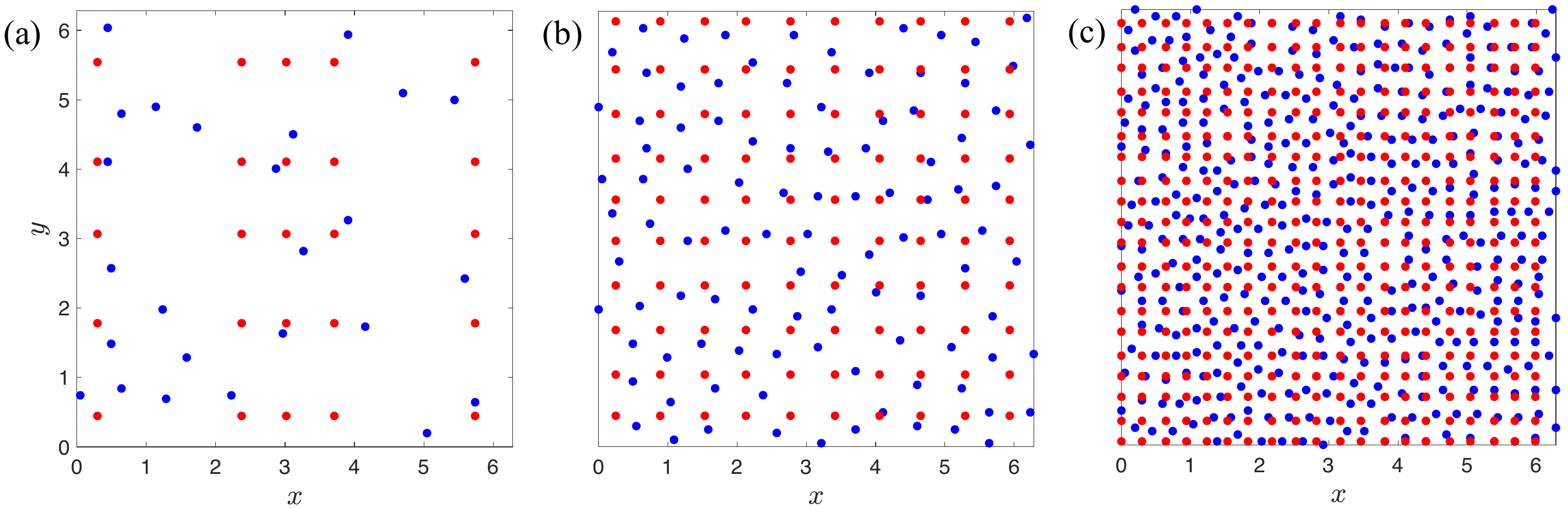}
    \caption{Optimal sensor locations obtained by vector-DEIM (blue circles) and tensor-DEIM (red circles) for the Kolmogorov flow. The number of sensors are (a) 25, (b) 100, and (c) 400 as in Fig.~\ref{fig:kolm_w}.}
    \label{fig:kolm_sensorLoc}
\end{figure}

\MF{We have verified that $10^3$ snapshots are adequate for the results to have converged. For instance, changing the number of snapshots to $800$ did not significantly alter the results reported below. Furthermore, choosing the training snapshots at random, instead of the first $75\%$, did not affect the reported results.}

Figure~\ref{fig:kolm_w} compares reconstruction results using the conventional vector-DEIM and our tensor-based DEIM. These reconstructions are performed for a vorticity field in the testing data set. As the number of sensors increases, both reconstructions improve. For the same number of sensors, our tensor-based method always returns a more accurate reconstruction compared to vector-DEIM. Furthermore, the storage cost for tensor-DEIM is much lower. For instance, for reconstruction from 400 sensors, storing the tensor bases  only requires $0.07\%$ of the memory required by vector-DEIM. \MF{Figure~\ref{fig:kolm_sensorLoc} compares the location of sensors used in vector-DEIM and tensor-DEIM. For relatively small number of 25 sensors, the optimal sensor locations corresponding to vector-DEIM and tensor-DEIM are significantly different. However, as the number of sensors increases the difference diminishes.}
\begin{figure}
    \centering
    \includegraphics[width=0.5\textwidth]{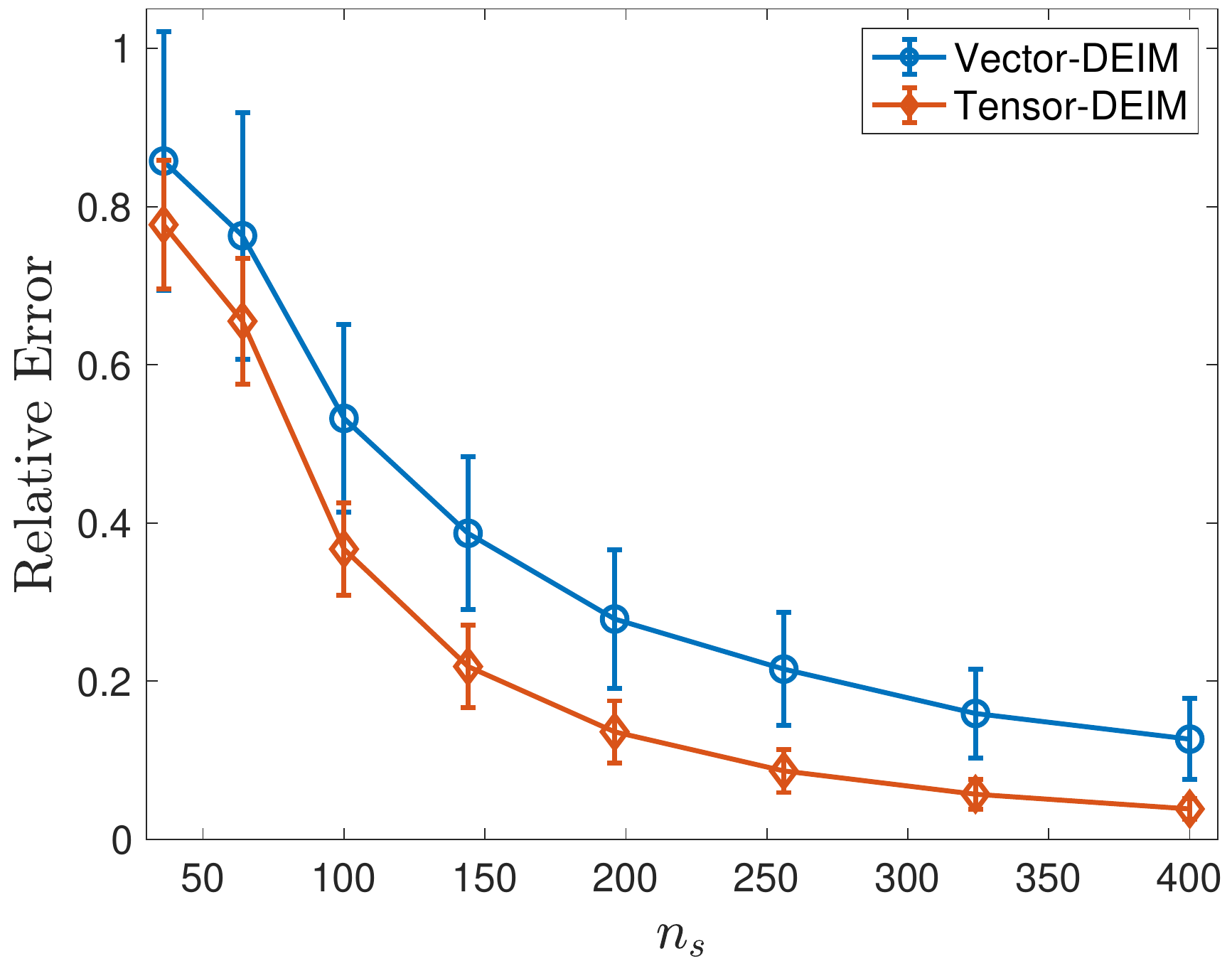}
    \caption{Relative flow reconstruction error for the Kolmogorov flow. The number of sensors is denoted by $n_s$.}
    \label{fig:rel_err}
\end{figure}

Figure~\ref{fig:rel_err} shows the relative reconstruction error as a function of the number of sensors. For each snapshot in the testing data set, we compute this error separately. The symbols in Fig.~\ref{fig:rel_err} mark the mean relative error taken over the 250 snapshots in this data set. The error bars show one standard deviation of the error. In every case, tensor-DEIM outperforms its vectorized counterpart as assessed by the mean relative error. In addition, the standard deviation of the relative error is smaller when using tensor-DEIM as compared to vector-DEIM.

\subsection{Sea surface temperature}
As the second test case, we consider the reconstruction of global
ocean surface temperature. The dataset is publicly available at NOAA Optimum Interpolation SST V2, noaa.oisst.v2~\citep{reynolds2002improved}. The temperature distribution is affected by the complex ocean flow dynamics resulting in seasonal variations. This dataset is in the form of a time series in which a snapshot is recorded every week in the span of 1990-2016 and data is available at a resolution of $1^\degree \times 1^\degree$. In total there are $1688$ snapshots, which we split into a training set of $1200$ (roughly $71\%$) and a testing set of $488$ snapshots. 
\begin{figure}
    \centering
    \includegraphics[scale=0.45]{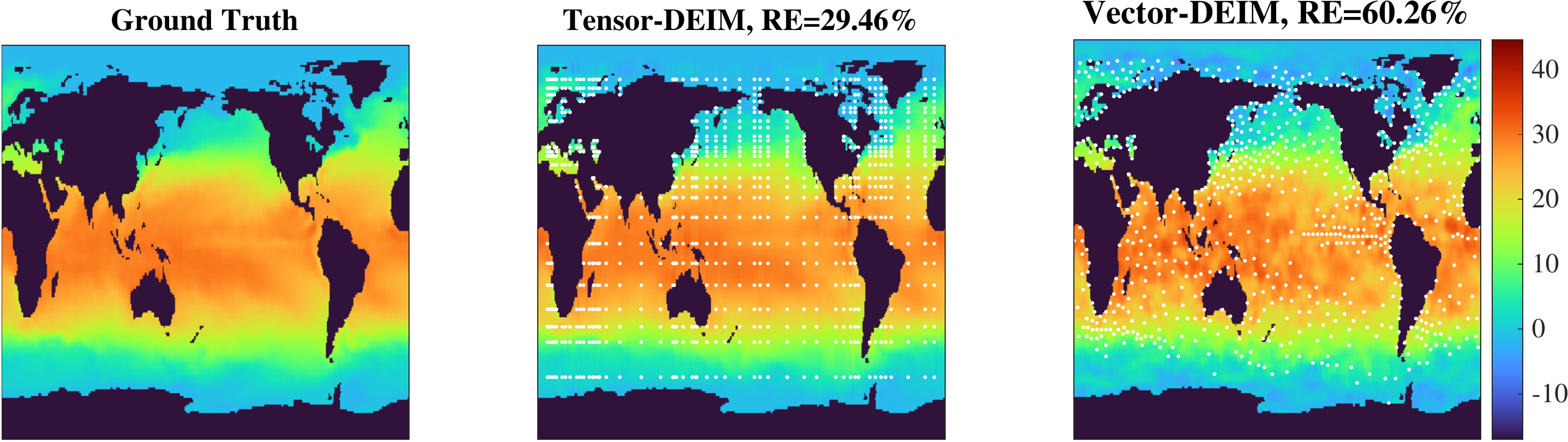}
    \caption{Comparing vector-DEIM and tensor-DEIM on the sea surface temperature dataset on March 3, 2013. The white dots indicate sensor locations and RE represents relative error. The color bar represents temperature in degrees centigrade.} 
    \label{fig:sst_plot}
\end{figure}

Figure~\ref{fig:sst_plot} shows the reconstruction using tensor-DEIM and vector-DEIM compared to the ground truth. The tensor-DEIM places sensors in a rectangular array and some sensors may fall within the land surface. These sensors are discarded, and only the ones on the ocean surface are retained. In this problem instance, there are $764$ sensors. To measure the accuracy of the reconstruction, we use the relative error of the fields centered around the mean sea surface temperature. As is seen from the figure, the reconstruction error from tensor-DEIM is superior to that of vector-DEIM. Storing the tensor bases  only requires $0.04\%$ of the memory footprint required by vector-DEIM.

\begin{figure}
    \centering
    \includegraphics[width=\textwidth]{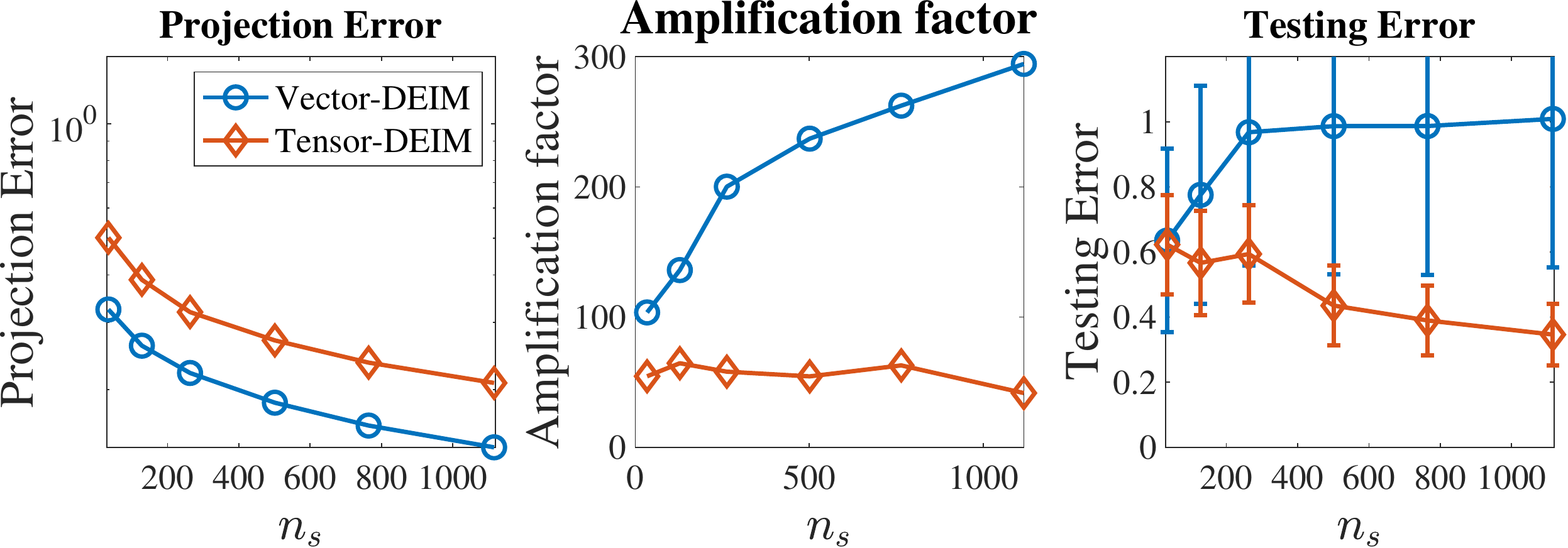}
    \caption{Comparing the contributions to the error in vector-DEIM and tensor-DEIM on a training set with first $71\%$ of the data. (left) Projection error normalized by the norm of the snapshot (center) Amplification factor, (right) Overall relative error. }
    \label{fig:interp_error}
\end{figure}

Figure~\ref{fig:interp_error} shows the relative error as a function of the number of sensors. As in the previous experiment, we compute the error over each snapshot in the test dataset and display the mean over the $488$ snapshots with the error bars indicating one standard deviation of the error. As can be seen, once again tensor-DEIM outperforms its vectorized counterpart both in terms of having a lower mean and standard deviation. The superiority of tensor-DEIM becomes more and more pronounced as the number of sensors increases. Note that the error in both methods does not decrease monotonically with more sensors. This is because the error has two contributions: the approximation of the snapshot by the basis and the amplification factor due to the interpolation; see Eq.~\eqref{eq:err_tensor}. While the first contribution is non-increasing, the second contribution may increase with an increasing number of sensors; hence the overall error may increase. 

\AS{To explain why the interpolation error for vector-DEIM gets worse with an increasing number of sensors, we plot the two different contributions to the error. By Lemma 3.2 of~\cite{Sorensen2010}, the testing error of vector-DEIM is bounded as 
\begin{equation} \| \V{f} -\M{U}_r (\M{S}^\top\M{U}_r)^{-1}\M{S}^\top\V{f} \|_2 \leq \| (\M{S}^\top\M{U}_r)^{-1}\|_2 \|\V{f} -\M{U}_r \M{U}_r^\top\V{f}\|_2.  \end{equation}
The error bound has two contributions: $\|\V{f} -\M{U}_r \M{U}_r^\top\V{f}\|_2$ which we call the projection error,  and  $\| (\M{S}^\top\M{U}_r)^{-1}\|_2$ which we call as the amplification factor. A similar classification can be done for the error due to tensor-DEIM, using~\eqref{eqn:tdeim_test_err}. The left panel of Fig.~\ref{fig:interp_error} shows the projection error normalized by $\|\V{f}\|_2$ and the center panel shows the amplification factor. We see that while the projection error decreases with an increasing number of sensors, the amplification factor increases, resulting in an overall increased error on average. On the other hand, for tensor-DEIM the projection error is higher compared to vector-DEIM. However, the amplification factor is nearly constant, and the overall error for tensor-DEIM decreases with an increasing number of sensors. }

\begin{figure}
    \centering
    \includegraphics[width=\textwidth]{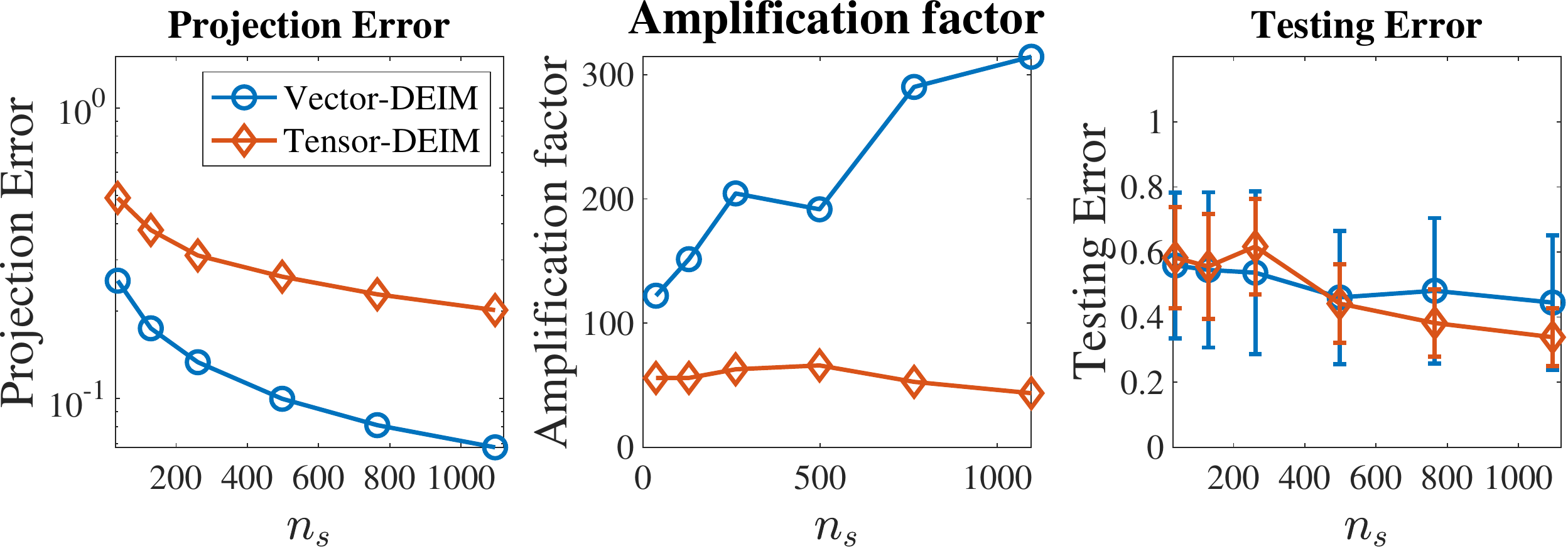}
    \caption{Comparing the contributions to the error in vector-DEIM and tensor-DEIM on a randomly selected training set. (left) Projection error normalized by the norm of the snapshot (center) Amplification factor, (right) Overall relative error. }
    \label{fig:interp_error_rand}
\end{figure}

\AS{To explore this further, we repeat the experiment but with a randomly generated training and testing split; more precisely, we randomly choose $\sim 71\%$ of the data to be the training set and the remaining to be the test set. The results are shown in Fig.~\ref{fig:interp_error_rand}.
Similar to Fig.~\ref{fig:interp_error}, we plot the contributions to the error in the left and center panels and the overall error in the right panel. Qualitatively, we see a similar trend as in the previous experiment. However, one major difference is that the overall mean error of vector-DEIM is now much closer to that of tensor-DEIM. But tensor-DEIM still has a lower mean error and a much smaller standard deviation.}

\MF{Note that this is in contrast with the Kolmogorov flow data, where randomization had no significant effect on the results. We attribute this to the fact that the Kolmogorov data contains a large and well-separated set of snapshots. Hence, the splitting of the data into training and test subsets does not play a major role on the sampling of the attractor.}

\begin{figure}[h!]
    \centering
    \includegraphics[width=\textwidth]{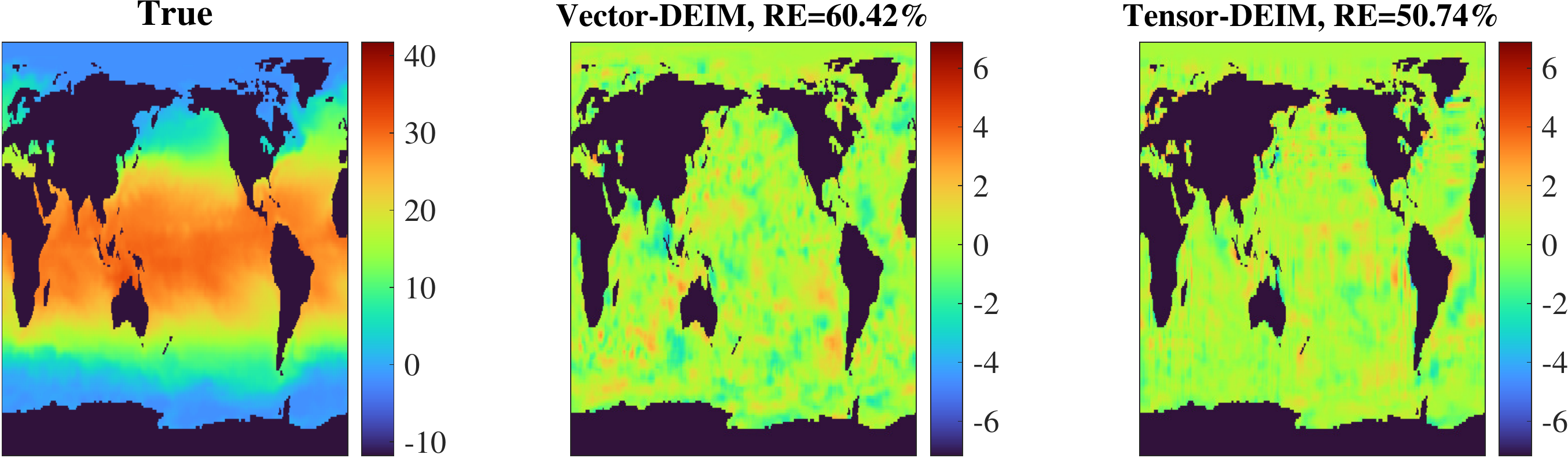}
    \caption{\AS{Comparing the spatial distributions of the error. Left: true SST from December 1997. Center: spatial distribution of error (reconstructed SST minus true SST) for vector-DEIM. Right: spatial distribution of error for tensor-DEIM.}}
    \label{fig:elnino}
\end{figure}

\MF{Finally, we turn to the problem of El Ni\~no-Souther Oscillation (ENSO), i.e., cycles of warm (El Ni\~no) and cold (La Ni\~na) water temperature in the Pacific ocean near the equator. It is known that reconstructing ENSO features from sparse measurements is challenging~\citep{Manohar2018,Maulik2020}. Here, we focus on the El Ni\~no event of the winter of 1997-1998, which is well-known for its intensity off the coast of Peru. In particular, Fig.~\ref{fig:elnino} (left panel) shows the SST in December 1997. This snapshot lies in the test dataset when the sets are chosen at random as opposed to sequentially. We see that the overall reconstruction error using vector-DEIM is larger compared to tensor-DEIM. More importantly, the spatial error near the El Ni\~no oscillation is significantly larger when using vector-DEIM. Therefore, tensor-DEIM more successfully reconstructs the El Ni\~no patterns. This is quite counter-intuitive since vector-DEIM tends to place more sensors in the El Ni\~no region (see the right panel of figure~\ref{fig:sst_plot} off the coast of Peru).}
\subsection{Three-dimensional unsteady flow} \label{sec:3D}
We consider a dataset obtained by the  simulation of an incompressible 3D flow around a CAD model of the research vessel Tangaroa (\cite{Popinet04:Tangaroa}). This dataset is obtained by large-eddy simulation using the Gerris flow solver~\citep{gerrisflowsolver}. The flow variables are resampled onto a regular grid in the spatial region of interest, $ [-0.35, 0.65] \times [-0.3, 0.3] \times [-0.5, -0.3]$. The reported spatial variables are dimensionless, normalized with the characteristic length scale $L=276\,$m, four times the ship length. Flow velocity is normalized by the inflow velocity $U=1\,$ m/s. We consider only the $u$ component of the velocity $(u,v,w)$ and subsampled the data to consider a grid size $150\times 90\times 60$. We split the available $201$ snapshots into a training dataset with $150$ snapshots ($\sim 75\%$) and a test dataset with $61$ snapshots.
\begin{figure}
    \centering
    \includegraphics[width=\textwidth]{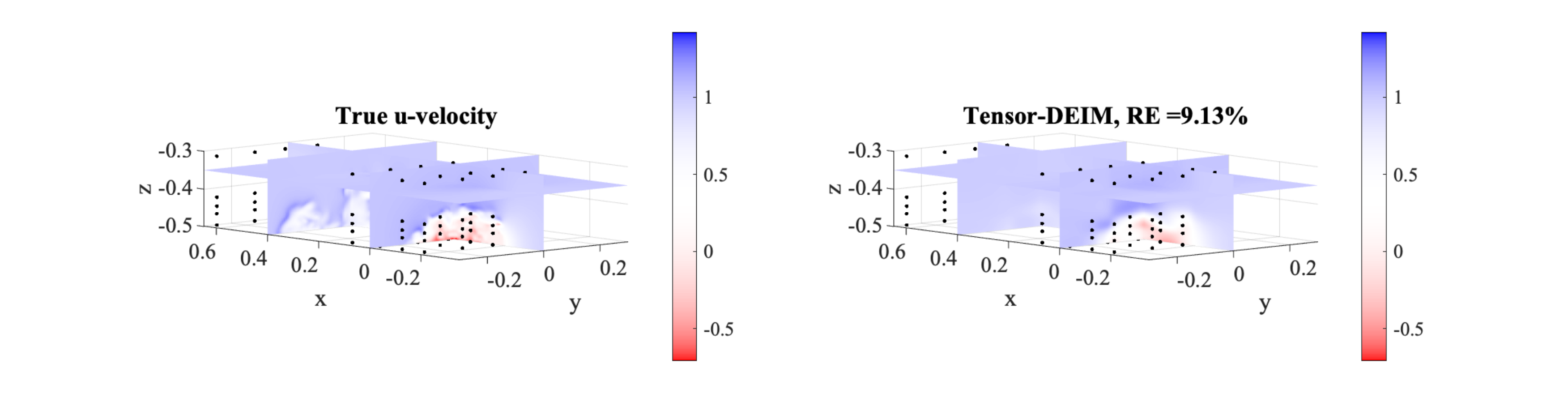}
    \caption{(left) The true u-velocity from the test dataset. (right) Reconstruction using tensor-DEIM with $125$ sensors. Black dots indicate the position of the sensors. RE denotes the relative error.}
    \label{fig:tangaroa}
\end{figure}

Due to the enormous size of the resulting tensor, we used randomized SVD to compute the factor matrices $\M\Phi_n$ for $1 \leq n \leq 3$ via the MATLAB command \verb|svdsketch|. We choose $5\times 5 \times 5 = 125$ sensors to reconstruct the flow field. The true $u$-velocity for a snapshot in the test dataset is plotted in the left panel of Fig.~\ref{fig:tangaroa}. The reconstruction using tensor-DEIM is displayed in the right panel of the same figure. The relative error in the reconstruction is around $9\%$ suggesting that tensor-DEIM is adequately reconstructing the snapshot. The corresponding error for the same number of sensors using vector-DEIM is $20.08\%$. As in the previous examples, it is seen that tensor-DEIM is far more accurate than vector-DEIM for the same number of sensors. The average relative error over the entire test data is $9.07\%$ for tensor-DEIM and $41.79\%$ for vector-DEIM. For some snapshots, the error in vector-DEIM is as high as $75\%$ and the error appears to increase for the later snapshots. Finally, the cost of storing the tensor-DEIM factor matrices is $0.0015\%$ of the cost of storing the vector-DEIM basis. Once again, tensor-DEIM proves to be more accurate and far more storage efficient compared to vector-DEIM. 

\section{Conclusions}
Our results make a strong case for using tensor-based methods for sensor placement and flow reconstruction. In particular, our tensor-based method significantly increases the reconstruction accuracy and reduces its storage cost. Our numerical examples show that the relative reconstruction errors are comparable to vectorized methods when the number of sensors is small. However, as the number of sensors increases, our tensor-based method is 2-3 times more accurate than its vectorized counterpart. The improvements in terms of the storage cost are even more striking: tensor-DEIM requires only \AS{$0.0015-0.07\%$} of the memory required by vector-DEIM. 

\AS{Future work could include the application of the method to reacting flows.} In such flows, the multidimensional nature of our tensor-based method allows for separate optimal sensor placement for the flow field and each chemical species. Although our method is a tensorized version of DEIM, a similar tensor-based approach can be applied to other flow reconstruction methods such as gappy POD, sparsity-promoting methods, and autoencoders.

\section*{Acknowledgements}
{The authors would like to acknowledge the NOAA Optimum Interpolation (OI) SST V2 data provided by the NOAA PSL, Boulder, Colorado, USA, from their website at \url{https://psl.noaa.gov}. \MF{We would also like to thank the Computer Graphics Lab (ETH Zurich) for making the 3D unsteady flow data used in section~\ref{sec:3D} publicly available.}}

\section*{Funding}
{This work was supported, in part, by the National Science Foundation through the awards DMS-1821149 and DMS-1745654, and Department of Energy through the award DE-SC0023188.}

\appendix
\section{Proof of Theorem~\ref{thm:mainten}}\label{sec:app}
We will need the following notation for the proof. We express $ \T{Y} = \T{G} \times_1 \M{A}_1 \dots \times \M{A}_d $ in terms of unfoldings as $\T{Y}_{d} = \M{A}_d \M{Y}_{(d)} ( \M{A}_{d-1} \otimes \dots \otimes \M{A}_1)^\top$. 
Here we use $\otimes$ to denote the Kronecker product of two matrices.
 
 The proof is similar to~\cite[Proposition 1]{kirsten2022multilinear}. Using the properties of matrix-unfoldings and Kronecker products, we can write an equivalent expression for the error 
 \[\| \T{G} - \T{G} \times_{n=1}^d \M\Pi_n\|_F = \| (\M{I} - \otimes_{n=d}^1 \M\Pi_n) \M{G}_{(d+1)}^\top\|_F .  \] 
 From $\M\Pi_n\M\Phi_n\M\Phi_n^\top=\M\Phi_n\M\Phi_n^\top$, we get $(\M{I} - \otimes_{n=d}^1 \M\Pi_n)(\M{I} - \otimes_{n=d}^1 \M\Phi_n\M\Phi_n^\top) = \M{I} -\otimes_{n=d}^1 \M\Pi_n.$. 
 Using this result and submultiplicativity
 \[\begin{aligned} \|  (\M{I}- \otimes_{n=d}^1\M\Pi_n)\M{G}_{(d+1)}^\top\|_F = & \>\| (\M{I} - \otimes_{n=d}^1 \M\Pi_n)(\M{I} - \otimes_{n=d}^1 \M\Phi_n\M\Phi_n^\top) \M{G}_{(d+1)}^\top\|_F\\
 \leq & \> \| (\M{I}- \otimes_{n=d}^1 \M\Pi_n) \|_2\|   (\M{I} - \otimes_{n=d}^1\M\Phi_n\M\Phi_n^\top)\M{G}_{(d+1)}^\top\|_F .  \end{aligned}\]
Since $\M\Pi_n$ is an oblique projector, so are $\otimes_{j=1}^d \M\Pi_n$ and $\M{I} - \otimes_{j=1}^d \M\Pi_n$. By~\cite[Theorem 2.1]{szyld2006many}, which applies since $\otimes_{j=1}^d \M\Pi_n $ is neither zero nor the identity, $\|\M{I} - \otimes_{j=1}^d \M\Pi_n\|_2 = \|\otimes_{j=1}^d \M\Pi_n \|_2 = \prod_{n=1}^d \|\M\Pi_n\|_2$. In the last step, we have used the fact that the largest singular value of a Kronecker product is the product of the largest singular values. Therefore, 
\[\begin{aligned}
\|  (\M{I}- \otimes_{n=d}^1\M\Pi_n)\M{G}_{(d+1)}^\top\|_F\leq  & \> \left( \prod_{n=1}^d \|  \M\Pi_n\|_2 \right)  \|(\M{I} - \otimes_{n=d}^1\M\Phi_n\M\Phi_n^\top)\M{G}_{(d+1)}^\top\|_F \\
= & \>  \left( \prod_{n=1}^d \| (\M{S}_n^\top\M\Phi_n)^{-1}\|_2 \right) \|\T{G}  - \T{G} \times_{n=1}^d \M\Phi_n\M\Phi_n^\top\|_F.
 \end{aligned}, \]
 since $\M\Phi_n$ and $\M{S}_n$ have orthonormal columns, so $\|\M\Pi_n\|_2 = \|(\M{S}_n^\top\M\Phi_n)^{-1}\|_2$. The result follows from~\cite[Corollary 5.2]{vannieuwenhoven2012new}.


\end{document}